\documentclass[a4paper,11pt]{article}

\usepackage{amssymb}

\begin{document}

\title{The Hsu-Harn-Mu-Zhang-Zhu group key establishment protocol is insecure}
\author{Chris J Mitchell\\
  Information Security Group, Royal Holloway,
  University of London, UK
}

\date{15th March 2018 (v2)}
\maketitle

\begin{abstract}
A significant security vulnerability in a recently published group key
establishment protocol is described.  This vulnerability allows a malicious
insider to fraudulently establish a group key with an innocent victim, with the
key chosen by the attacker.  This shortcoming is sufficiently serious that the
protocol should not be used.
\end{abstract}

\section{Introduction}

Hsu, Harn, Mu, Zhang and Zhu \cite{Hsu17} recently published a protocol (which
for convenience we refer to as the HHMZZ protocol) designed to provide
authenticated group key establishment in a wireless network (although there are
no obvious wireless-specific aspects of the protocol).  In this brief note we
describe a serious security issue with this scheme; in particular it does not
provide the properties claimed.

The remainder of the paper is structured as follows.  Section~\ref{protocol}
sets out the protocol, including the intended context of use.
Section~\ref{analysis} then describes a serious security vulnerability in the
protocol.  The paper concludes in section~\ref{end}.

\section{The Hsu-Harn-Mu-Zhang-Zhu protocol} \label{protocol}

\subsection{Context and goals}  \label{goals}

The HHMZZ protocol is intended for use by a pre-established collection of
users, and enables any subset (group) of this community to be equipped with a
shared secret key by a trusted \emph{Key Generation Centre} (KGC)\@.  Such
protocols have been widely discussed in the literature --- see, for example,
chapter 6 of Boyd and Mathuria \cite{Boyd03}.  The area is sufficiently
well-established that an ISO/IEC standard for group key establishment
\cite{ISO11770-5:11} was published in 2011.

The threat model for such protocols varies, but typically the goal is that,
after completion of the protocol, all participants agree on the same key, they
know it is `fresh', and that no parties other than those intended learn
anything about the key. The authors  of the HHMZZ protocol are a little vague
as to the assumed capabilities of attackers, although they do refer to both
insider and outsider attackers (\cite{Hsu17}, section 4).  It is thus
legitimate to assume that the claimed protocol properties are intended to apply
even in a malicious insider scenario.

Note that the protocol is described in a way that seems to imply that the group
of users between which a key is established is always the same.  However,
closer examination, e.g.\ of the discussion following Theorem 3 of section 4,
reveals that this is a result of notational assumptions made to simplify the
presentation. To avoid confusion, in the description of the protocol given
below we have generalised the notation slightly to make it clear that the group
can change.

\subsection{Related work}  \label{related}

The HHMZZ protocol uses a combination of cryptographic hash-functions and a
secret sharing scheme.  The use of secret sharing as part of a group key
establishment protocol is long-established (see, for example, section 6.7.2 of
Boyd and Mathuria \cite{Boyd03}).  However, this approach is known to have
shortcomings, \cite{Boyd03}.

Indeed, the fact that the HHMZZ protocol has serious flaws is hardly surprising
given the unfortunate history of the area.  Back in 2010, Harn and Lin
\cite{Harn10} described a group key transfer protocol based on secret sharing
which is not only mathematically flawed, but also possesses very serious
security issues; not only did this give rise to a number of papers pointing out
the flaws (see, for example, \cite{Nam11,Nam12}), but also further flawed
protocols attempting to `fix' the flaws in the original scheme.  Some of the
rather sad history of the area can be found in the recent paper of Liu et al.\
\cite{Liu17}.

\subsection{The protocol}  \label{operation}

The following requirements apply for use of the protocol.  Note that we have
made some minor changes to the notation of Hsu et al.\ \cite{Hsu17} for the
purposes of clarity.
\begin{itemize}
\item The protocol is designed to work within a set of users ${\cal
    U}=\{U_i\}$, all of which must have registered with the KGC (and this
    KGC must be trusted by all users to generate and distribute secret
    keys).
\item All participants must agree on a large `safe' prime $p$ and a
    representation of the finite field $\mathbb{K}=\mathbb{Z}_p$ of $p$
    elements.  The participants must also agree on two cryptographic
    hash-functions $h_1$ and $h_2$, both mapping to $\mathbb{K}$.
\item All participants must agree on the function
    $\mathbf{v}_m:\mathbb{K}\rightarrow\mathbb{K}^{m+1}$ defined by:
    \[ \mathbf{v}_m(x) = (1,x,x^2,\ldots,x^m) \]
    (where $m\geq 2$).
\item Every user $U_i$ must:
     \begin{itemize}
     \item have a unique identifier $\mbox{ID}_i$;
     \item choose a secret key $x_i\in\mathbb{K}$, which is shared with
         the KGC.
     \end{itemize}
\end{itemize}

Now suppose an \emph{initiator} wishes to arrange for a new secret key to be
shared by the members of a group of users ${\cal U}'$ (${\cal U}'\subseteq{\cal
U}$). Suppose ${\cal U}' = \{U_{z_1},U_{z_2},\ldots,U_{z_t}\}$ for some $t\geq
2$.

The protocol proceeds as follows (where all arithmetic is computed in
$\mathbb{K}$).
\begin{enumerate}
\item The initiator sends a key generation request to the KGC along with
    the set of $t$ identifiers $\{\mbox{ID}_i: i\in{\cal U}'\}$.
\item The KGC broadcasts the set of identifiers $\{\mbox{ID}_i: i\in{\cal
    U}'\}$ as a response.
\item Each user $U_{z_j}$ in ${\cal U}'$, i.e.\ each user $U_{z_j}$ for
    which $\mbox{ID}_{z_j}$ is in the broadcast set of identifiers, chooses
    a fresh random challenge $r_j\in\mathbb{K}$ and sends it to the KGC.
\item The KGC performs the following steps.
\begin{enumerate}
\item The KGC randomly chooses a group key $S\in\mathbb{K}$ and a value
    $r_0\in\mathbb{K}$, and assembles the $(t+1)$-tuple
    $\mathbf{r}=(r_0,r_1,r_2,\ldots,r_t)$.
\item For every $i$ ($1\leq i\leq t$) the KGC now computes the inner
    product
    \[ s_{z_i} = ( \mathbf{v}_t(x_{z_i}+h_1(x_{z_i}||r_i||r_0)), \mathbf{r}) \]
    where $||$ denotes concatenation of bit strings (and the finite
    field values that are concatenated are converted to bit strings
    using an agreed representation).  The KGC also computes
    $u_{z_i}=S-s_{z_i}$.
\item The KGC now computes the tag \emph{Auth} as
\[
\mbox{\emph{Auth}}=h_2(S||\mbox{ID}_1||\mbox{ID}_2||\ldots||\mbox{ID}_t||r_0||r_1||r_2||\ldots||
r_t||u_{z_1}||u_{z_2}||\ldots||u_{z_t}) \]

where, as previously, in assembling the input to $h_2$, elements of
    $\mathbb{K}$ are converted to bit strings using an agreed
    representation.
\item Finally, the KGC broadcasts
    \[ \mbox{\emph{Auth}}, r_0, (u_{z_1},u_{z_2},\ldots,u_{z_t}) \]
to all members of the group ${\cal U}'$.
\end{enumerate}
\item On receipt of the broadcast, each user $U_{z_i}\in{\cal U}'$ ($1\leq
    i\leq t$) proceeds as follows.
\begin{enumerate}
\item $U_{z_i}$ computes
    \[ s_{z_i} = ( \mathbf{v}_t(x_{z_i}+h_1(x_{z_i}||r_i||r_0), \mathbf{r})) \]
using its secret key $x_{z_i}$, the random challenges $r_i$ ($1\leq
i\leq t$) sent earlier in the protocol, and the broadcast value $r_0$.
\item $U_{z_i}$ now computes the group key as $S=u_{z_i}+s_{z_i}$.
\item Finally, $U_{z_i}$ verifies the tag \emph{Auth} by recomputing it
    using the newly computed group key and the values sent earlier in
    the protocol.
\end{enumerate}
\end{enumerate}

\subsection{Two minor observations}

In the form that the protocol is specified by Hsu et al.\ \cite{Hsu17}, every
participating group member is required to intercept the random challenges $r_j$
sent by every other group member to the KGC\@.  This seems likely to be
problematic, at least in some environments.  It would make more sense for the
KGC to broadcast the values $(r_1,r_2,\ldots,r_t)$ to all group members in step
4d of the protocol.

It would appear that the computation $x_{z_i}+h_1(x_{z_i}||r_i||r_0)$ is
intended to be a one-way function of the three field elements $x_{z_i}$, $r_i$
and $r_0$.  However, if $h_1$ is chosen appropriately, precisely the same
property can be achieved without the addition, i.e.\ simply by computing
$h_1(x_{z_i}||r_i||r_0)$.

\subsection{Security claims} \label{claims}

Hsu et al.\ \cite{Hsu17} (see Theorem 1 of section 4) make the following claims
regarding the security of the protocol.  They state: `The proposed protocol
achieves the security features\footnote{In fact, the paper refers to `security
feathers', but this is presumably a misprint.} with key freshness, key
confidentiality and key authentication'.  In the `proof' of Theorem 1, the
following statements are made.
\begin{quote}
\emph{Key authentication} is provided through the value \emph{Auth} in step 4.
\ldots~~Any insider also cannot forge a group key without being detected since
the group key is a function of each member's long-term secret $x_i$.
\end{quote}
As we show below, this claim is incorrect; that is, an insider \emph{can} forge
a group key.

\section{Analysis} \label{analysis}

We now describe a serious security vulnerability in the protocol.

\subsection{Attack goal and model}

We consider the scenario where a `victim user' $U_v$ is a member of a group
${\cal U}'$ of $t$ users for which a new key is requested. We make the
following assumptions.

\begin{itemize}
\item One of the users, $U_m$ say, in the group ${\cal U}'$ is malicious.
\item $U_m$ can control the channel between the KGC and the victim user
    $U_v$.  In fact $U_m$ only needs to be able to modify the content of
    what $U_v$ receives in the final broadcast from the KGC sent in step
    4d.
\item $U_m$ wishes to make $U_v$ accept a key $S^*$ of the malicious user's
    choice.
\end{itemize}

\subsection{Attack operation}

The attack is very simple to describe.  We suppose that the protocol proceeds
as described in section~\ref{operation}, where $U_v, U_m\in{\cal U}'$.

In step 4d, $U_m$ intervenes and prevents the broadcast from the KGC reaching
$U_v$.  Because the malicious user is a valid member of ${\cal U}'$, $U_m$ can
calculate the secret key $S$ generated and distributed by the KGC\@. $U_m$ now
chooses a different secret key $S^*\in\mathbb{K}$, and computes
\[ u^*_v= u_v-S+S^* \]
and
\[
\mbox{\emph{Auth}}^*=h_2(S^*||\mbox{ID}_1\ldots||\mbox{ID}_t||r_0||r_1||\ldots||
r_t||u_1||\ldots||u_{v-1}||u^*_v||u_{v+1}||\ldots||u_t). \]

That is, \emph{Auth}$^*$ is computed using exactly the same inputs as
\emph{Auth} except that $S$ and $u_v$ are changed to $S^*$ and $u^*_v$.

$U_m$ now sends a modified version of the KGC's broadcast to $U_v$, where
\emph{Auth} and $u_v$ are replaced by \emph{Auth}$^*$ and $u^*_v$.  It is
straightforward to see that the victim user $U_v$ will compute the secret key
as $S^*$, and the tag \emph{Auth}$^*$ will verify correctly.  The attack is
complete.

\section{Conclusions}  \label{end}

As demonstrated above, the HHMZZ protocol fails to possess the properties
claimed of it.  This means that the protocol should not be used.  It is
important to observe that Hsu et al.\ \cite{Hsu17} do not provide a rigorous
security proof using state of the art `provable security' techniques, nor do
they give a formal model of security for the protocol. This helps to explain
why fundamental flaws exist.  Indeed, the following observation, made by Liu et
al.\ \cite{Liu17} with respect to a number of previously proposed but flawed
group key establishment protocols, is highly relevant.
\begin{quote}
The security proof for each vulnerable GKD protocol only relies on incomplete
or informal arguments.  It can be expected that they would suffer from attacks.
\end{quote}

It would, of course, be tempting to try to repair the protocol to address the
issues identified, but, unless a version can be devised with an accompanying
security proof, there is a strong chance that flaws will remain.

%\bibliographystyle{plain}
%\bibliography{crypto}

\end{document}